\documentclass[12pt]{article}
\title{RG solutions for $\alpha_s$ at large $N_c$ in  $d=3+1$ QCD}
\author{ Yu.A.Simonov\\
State Research Center\\Institute of Theoretical and Experimental
Physics, \\ Moscow, 117218 Russia}
\date{}

\newcommand{\beq}{\begin{eqnarray}}
 \newcommand{\eeq}{\end{eqnarray}}
\newcommand{\be}{\begin{equation}}
 \newcommand{\ee}{\end{equation}}

\def\fun#1#2{\lower3.6pt\vbox{\baselineskip0pt\lineskip.9pt
\ialign{$\mathsurround=0pt#1\hfil ##\hfil$\crcr#2\crcr\sim\crcr}}}

\newcommand{{\SD}}{\rm SD}

\newcommand{\lan}{\langle}
\newcommand{\ran}{\rangle}

\begin{document}

\maketitle

\begin{abstract}
Solutions of RG equations for   $\beta(\alpha)$ and $\alpha(Q)$
are found in the class of meromorphic functions satisfying
asymptotic conditions at large  $Q$ (resp. small $\alpha)$, and
analyticity properties  in the  $Q^2$ plane. The resulting
$\alpha_R(Q)$ is finite in the Euclidean $Q^2$ region and agrees
well at $Q\geq 1$ GeV
 with the $\overline{MS}~~ \alpha_s(Q)$.
 \end{abstract}

 \section{Introduction}

 QCD is known to simplify at large $N_c$:
 i) the perturbation theory dictates that only planar  diagrams
 survive in the leading order \cite{1}; ii)  assuming that
 confinement exists also in the limit $N_c\to \infty$, the
 spectrum of mesons and glueballs consists of bound-state poles,
 and the decay width vanishes at large $N_c$\cite{1,2,3}.

 In the $1+1$ QCD this property was proved both analytically and
 numerically \cite{3} (for a review see \cite{4})  since confinement
 in this case is induced by the perturbative gluon exchange.

 On the lattice numerical calculations have confirmed that the
 large $N_c$ limit is achieved with few percent accuracy already
 at $N_c=3$ for several tested observables \cite{5}. It is thus
 likely that the large $N_c$ QCD is a good first approximation for
 the realistic $N_c=3$ case, which has an accuracy of 10\% or
 better.

 From theoretical point of view the large $N_c$ limit of QCD is
 very useful since it drastically simplifies the analytic
 structure of amplitudes, e.g. one has sums of simple poles in the
 two-point function, and for the four-point function one expects
 formulas of the Veneziano type.

 Thus one expects the large $N_c$ amplitudes as functions of
 external momenta to be meromorphic. On the other hand the
 perturbation series yields the typical logarithmic dependencies
 and the RG equation prescribes for $\alpha_s(Q)$ the  structure
 with unphysical poles and cuts which are incompatible with
 unitarity and analyticity. E.g. the one-loop expression for
 $\alpha_s (Q)$ has the form
 \be
 \alpha_s (Q)=\frac{4\pi}{\beta_0 \ln Q^2/\Lambda^2},~~~\beta_0
 =\frac{11}{3} N_c -\frac23 n_f\label{1}\ee
 with the Landau ghost pole at $Q^2=\Lambda^2$ and the cut with
 the branch point $Q^2=0$. This property being true for all $N_c$,
 strikingly violates the expected meromorphic structure of
 amplitudes for $N_c\to \infty$.

 One might argue that $\alpha_s(Q)$ by itself is not yet the
 physical amplitude, and in the latter the unphysical features of
 $\alpha_s$ may be somehow compensated. This is however
 contradicted by examples of amplitudes, e.g. for the process
 $e^+e^-\to$ hadrons, where $\frac{\alpha_s(Q)}{\pi}$ enters
 directly into the hadronic ratio $R(s)$.\be
 R(s)=R_{parton}
 +\frac{\alpha_s(s)}{\pi}+O\left(\frac{\alpha_s}{\pi}\right)^2.\label{2}\ee
 Moreover one can define the "effective coupling" $\alpha_R(Q)$
 for the process $R$, which enters the physical  amplitude
 $\Omega_R$ directly \cite{6}
 \be \Omega_R=\Omega_R^{(0)} +\omega_R\alpha_R(Q)\label{3}\ee
 and this "process dependent" $\alpha_R$ satisfies the standard
 RG equation
 \be
 \frac{d\alpha_R(\mu)}{d\ln \mu}=
 \beta^{(R)}(\alpha)=-\frac{\beta_0^{(R)}\alpha^2}{2\pi}-\frac{\beta_1^{(R)}\alpha^3}{4\pi^2}+O(\alpha^4)\label{4}\ee
 where $\beta^{(R)}_0$ and $\beta_1^{(R)}$ are standard
 scheme-independent  coefficients, while $\beta_n^{(R)}, n>1$
 depend on the process. One can easily  see, that the solution of
 (\ref{4}) has the dominant asymptotic term of the form (\ref{1})
 with the unphysical features discussed above.

 One might still argue that perturbative series and $\alpha_s(Q)$
 itself should be considered only in the asymptotic regime when
 $Q$ is large and therefore the logarithmic singularities and
 Landau ghost pole are far away.

 However the analytic structure of Riemann surfaces with cuts is
 different everywhere from that of meromorphic function and this
 argument of asymptotics can be turned around to imply that the
 logarithmic asymptotics of $\alpha_s(Q)$ is an asymptotic
 approximation of the true analytic function which is meromorphic
 at large $N_c$ and is in agreement with all expected properties
 of large $N_c$ physical amplitudes.

 It is the purpose of the present paper to exemplify the solutions
 of the RG equations which have the desired meromorphic
 properties. These solutions will have the standard logarithmic
 asymptotics in good numerical and analytic similarity with the
 standard perturbation theory.

 Moreover we show that this meromorphic-logarithmic duality has
 deeper roots in the quark-hadron duality and provide explicit
 example of this connection.

 The paper is organized as follows. In section 2 the RG equation
 for $\alpha_s(Q)$ is considered and the general form of solution
 is written down, containing an arbitrary function with known
 asymptotics. In section 3 the meromorphic solution is suggested
 and its properties are studies. In section 4 connection with the
 standard perturbation theory is investigated. The concluding
 section is devoted to general discussion of the meromorphic --
 logarithmic duality -- in QCD and QED.

 \section{General solutions of RG equations}

 In this section we follow the line of reasoning which was given
 in \cite{7}. We write the RG equation as in (\ref{4})  suppressing the
 subscript $R$ everywhere and express $\beta(\alpha)$ through
 another unknown function $\varphi\left( \frac{1}{\alpha}\right)$:
 \be
 \frac{d\alpha(\mu)}{d\ln \mu} =\beta(\alpha), ~~ \beta(\alpha)
 =-\frac{\beta_0}{2\pi}
 \frac{\alpha^2}{\left[1-\frac{\beta_1}{2\pi\beta_0}\varphi'\left(\frac{1}{\alpha}\right)\right]}.\label{5}\ee

Here $\varphi'(x)$ means the  derivative of function $\varphi(x)$
in the argument $x$, and $\mu$ is the RG scale, which will be
later traded as usual for external parameters (momenta) of the
given process $P^2_i$ since $\alpha$ can depend only on the ratio
$P_i^2/\mu$.

In terms of the function $\varphi$ the solution of the RG equation
for $\alpha(\mu)$ can be immediately written
\be
\alpha=\frac{4\pi}{\beta_0\left[\ln \mu^2+\chi+
\frac{2\beta_1}{\beta_0^2}\varphi\left(\frac{1}{\alpha}\right)\right]}.\label{6}\ee
Here $\chi$ is an arbitrary function of $P_i$ such that
$\ln\mu^2+\chi$ is some function of $P_i^2/\mu^2, P^2_i/P^2_j$ the
form of which depends on the process and will be found below for
concrete examples.

The Standard Perturbation Theory (SPT) which is assumed to be
valid at large  $Q^2$, provides some limitations on the properties
of  functions $\chi$ and $\varphi$, which one must impose. Namely,
the SPT Taylor expansion of $\beta(\alpha)$ is known for the first
four terms
\be
\beta(\alpha)= -\frac{\beta_0}{2\pi} \alpha^2
-\frac{\beta'}{4\pi^2} \alpha^3-\frac{\beta_2}{64\pi^3}\alpha^4
-\frac{\beta_3}{(4\pi)^4} \alpha^5 +O(\alpha^6).\label{7}\ee Here
$\beta_0, \beta_1$ are scheme-independent and for $n_f=0$ are
equal to $\beta_0 =\frac{11}{3} N_c-\frac23 n_f$ and
$\beta_1=\frac{17}{3} N_c^2-\frac53 N_c n_f - N_c  n_f,$ while
$\beta_2$ and $\beta_3$ have been calculated in the
$\overline{MS}$ scheme. Comparing (\ref{7}) and (\ref{5}) one
obtains the following condition on the asymptotic behaviour of
$\varphi(x)$ at large $x (x\equiv \frac{1}{\alpha})$: \be
\left.\varphi(x)\right|_{x\to\infty}\cong \ln x
+O\left(\frac{1}{x}\right).\label{8}\ee The first term on the
r.h.s. of (\ref{8}) reproduces the scheme-independent coefficients
$\beta_0, \beta_1$ in the expansion (\ref{7}), while the term
$O(\frac{1}{x})$ contributes to the higher order coefficients
$\beta_2, \beta_3,...
$
Conditions on  the function $\chi(P^2)$ are more subtle and in
general depend on the process in question. Here one should
distinguish in $P^2_i$ the external parameters which can be made
arbitrarily large, as the momentum $Q^2$ in the two-point function
$\Pi(Q^2)$ or in the  formfactor, and other renorm-invariant
parameters, which define the scale of confinement, e.g. the string
tension $\sigma$, or the RG invariant gluonic condensate
$\frac{\beta(\alpha)}{16\alpha} \lan Fâ_{\mu\nu} (0)
F_{\mu\nu}â(0)\ran$. In the framework of the Background
Perturbation Theory (BPT) \cite{8,9} both appear  as vacuum
expectation values of operators made of RG invariant combinations
$gF_{\mu\nu}$, where $F_{\mu\nu}$ refers to the background field.
Here we do not use BPT, but only consider all RG invariant
parameters like $\sigma$ on the same ground as the true external
parameters -- the  external momenta like $Q^2$. In what follows we
shall use the generic mass parameter $m^2\equiv 4\pi\sigma$ as the
confinement scale and disregard for simplicity other vacuum field
characteristics, like the gluonic correlation length $\lambda$,
which can be computed in principle and in practice \cite{10}
through $m^2$.

As a result keeping only one external momentum $Q^2$, $\chi$ can
be written as $\chi\equiv \chi(Q^2/m^2)$. Then the large $Q^2$
behaviour of $\alpha(Q^2)$, \be \alpha(Q^2)(Q^2\to\infty) \sim
\frac{4\pi}{\beta_0\ln Q^2/\Lambda^2}+O\left(\frac{\ln \ln
Q^2}{(\ln Q^2)^2}\right)\label{9}\ee dictates the  following
behaviour of $\chi+\ln \mu^2$, where we go over to the standard
$\Lambda_{QCD}$ parametrization instead of $\mu$
parametrization\be(\chi+\ln \mu^2)\equiv\ln \frac{m^2}{\Lambda^2}
+\Psi(Q^2, m^2),~~ \Psi|_{Q\to\infty}\cong\ln
\frac{Q^2}{m^2}.\label{10}\ee

\section{Meromorphic realization of RG solutions}

As was discussed  in Introduction, at large $N_c$ the analytic
structure of physical  amplitudes and  of  effective charge
$\alpha(Q)$ is simplified and reduced to the meromorphic functions
with simple poles at $Q^2=-M^2_n$, corresponding to the bound
states of quarks and gluons -- mesons, glueballs and hybrids.

Therefore the function $\Psi ( Q^2, m^2)$ in (\ref{19}) can be
represented in the spectral form
\be
\Psi (Q^2, m^2) =-\sum^\infty_{n=0} \frac{c_n}{Q^2+M^2_n} + const
\label{11}\ee where the coefficients $c_n$ are independent of
$Q^2$. The spectrum $M^2_n$ was found for large $N_c$  for mesons,
glueballs and hybrids in the limit of  no mixing between them
\cite{11}. In the large $N_c$ limit the meson-glueball mixing
vanishes,  while the  meson-hybrid mixing is $O(\alpha)$
\cite{12},  and following \cite{11} we write  the lowest order
 in $N_c$ and $\alpha$-independent spectrum, for glueballs and
mesons as \be M^{(0)2}_n=m^2(n+L/2) +M^2_0 +O(1/n)\label{12}\ee
where $m^2= 4\pi \sigma$, and for glueballs $\sigma(glue) =\frac94
\sigma (quark)$. In what follows we  shall be interested only in
the asymptotic part of the spectrum at large $n$ and $L=0,1$, (the
corrections to (\ref{12}) for $n=1,2$ come from spin splittings
and is of the order of 20\%). In general one can rewrite the
effective coupling for the given process $R$ with external
parameters denoted as $Q^2$ as
\be
\alpha_R = \frac{4\pi}{\beta_0\left [\ln \frac{m^2_R}{\Lambda^2}+
\Psi_R (Q^2, m^2_R) + \frac{2\beta_1}{\beta^2_0} \varphi_R
\left(\frac{1}{\alpha_R}\right) \right]}.\label{12a}\ee Here
$m^2_R, \Psi_R$ and $\varphi_R$ depend on the process and $\Psi_R$
contains the poles of  the spectrum in the lowest (one loop)
approximation. Conditions on $\Psi_R, \varphi_R$, Eqs. (\ref{10})
and (\ref{8}) respectively, impose restrictions on coefficients
$c_n, M_n^{(0)}$ in the spectral representation (\ref{11}). As an
example one can choose $\Psi_R(Q^2, m^2_R)$ in the following form
\be
\Psi_R\equiv\psi\left(\frac{Q^2+(M^{(0)}_R)^2}{m^2_R}\right),~~
\psi(x) =\frac{\Gamma'(x)}{\Gamma(x)}\label{13}\ee which implies
that the spectrum responsible for the one-loop RG evolution of
$\alpha_R(Q)$ is \be M^2_{RN} (1-{\rm loop})=m^2_R n+
(M^{(0)}_R)^2.\label{14}\ee

The asymptotics of $\psi(x)$ at large $x$ is \be \psi(x) =\ln x
-\frac{1}{2x} -\sum^\infty_{k=1} \frac{B_{2k}}{2k
x^{2k}}\label{15}\ee where $B_{2k}$ are Bernoulli numbers behaving
at large $k$ as $B_{2k}\sim 2(-)^{k-1} (2k)!/(2\pi)^{2k}$.

Therefore the asymptotic condition (\ref{10}) is  satisfied by the
choice (\ref{13}). Note that one can also satisfy this condition
by the realistic spectrum which asymptotically has the form
(\ref{14}) but differs from it for the first $N_0$ terms, provided
the coefficients $c_n$ tend to constant for large $n$.

At this point it is important to  define the physical system which
provides the one-loop spectrum (\ref{14}). To this end one  can
compare
 the one-loop expansion in SPT, $\alpha_s(1-{\rm
 loop})=\alpha^{(0)}_s-\frac{\beta_0}{4\pi}  \ln( Q^2/\mu_0^2)
 (\alpha_s^{(0)})^2+...$ with the equivalent expansion of
 $\alpha_R$ (\ref{12a}):
 $$\alpha_R("1-{\rm
 loop}")=\alpha^{(0)}_R-\frac{\beta_0}{4\pi} \Psi_R (Q^2, m^2_R)
 (\alpha_R^{(0)})^2+...$$
 where $\alpha^{(0)}_R =\frac{4\pi}{\beta_0\ln
 \frac{m^2_R}{\Lambda^2}}$. One can see that $\Psi_R(Q^2, m^2_R)$
 plays the role of gluon loop, or better, two-gluon intermediate
 states imbedded in the framework of the process $R$. In
 particular, when $R$ is the $e^+e^-$ annihilation into hadrons,
 then $\Psi_R$ is responsible for bound states of quark, antiquark
 and two gluons, i.e. in the large $N_c$ limit this is the
 two-gluon hybrid state.

 \section{Meromorphic solution and Standard  Perturbation Theory}

 In previous section it was argued that the general solution of
 RG equation (\ref{5}) which correctly reproduces the one-loop
 result of SPT and scheme-independent coefficients $\beta_0,
 \beta_1$ in the expansion of $\beta(\alpha)$ is given by

\be
\alpha_R = \frac{4\pi}{\beta_0\left [\ln \frac{m^2_R}{\Lambda^2}+
\psi \left(\frac{Q^2+ (M_r^{(0)})^2}{m^2_R}\right)  +
\frac{2\beta_1}{\beta^2_0} \varphi_R
\left(\frac{1}{\alpha_R}\right) \right]}.\label{16}\ee where
$\varphi_R(x)\to \ln x,  x\to \infty$. The  analytic properties of
$\alpha_R(Q^2)$ depend on the form of
$\varphi_R\left(\frac{1}{\alpha}\right)$. Neglecting the latter
(one-loop approximation)  one obtains for $\alpha_R$ the
meromorphic function of $Q^2$ with  simple poles at the negative
values of $Q^2$ (for $m^2_R>\Lambda^2$). If one adopts
$\varphi_R(x)=\ln x$, then one can reproduce the two-loop result
of SPT, but then $\alpha_R$ acquires  the logarithmic branch
points, in the $Q^2$ plane which are unacceptable for $N_c\to
\infty$. To keep the correct meromorphic properties of $\alpha_R$
one can choose $\varphi_R\left(\frac{1}{\alpha_R}\right)$ as a
meromorphic function ot its argument. A particular choice  was
made in \cite{7}, namely
\be
\varphi_R\left(\frac{1}{\alpha_R}\right) = \psi\left(
\frac{4\pi}{\beta_0\alpha_R}+\Delta\right) \label{17}\ee where
$\psi(x)$ is the same as in (\ref{13}) with the asymptotics
(\ref{15}) which satisfies asymptotic condition on $\varphi_R(x)$
given above. The resulting expression for $\beta(\alpha)$ is
\be
\beta(\alpha)=-\frac{\beta_0}{2\pi}
\frac{\alpha^2}{1-\frac{2\beta_1}{\beta_0}\psi'\left(\frac{4\pi}{\beta_0\alpha}
+\Delta\right)}.\label{18}\ee One can persuade oneself that for
$\Delta>\Delta_0 =1.255,~ \beta(\alpha)$ is analytic for
$\alpha>0$, and has infinite number of zeros for negative $\alpha$
condensing on the negative side of the point $\alpha=0$.  For
positive $\alpha$ the function $\beta(\alpha)$ is always negative
and monotonically decreasing, and thus has no IR fixed point. Also
the coefficients  $\beta_2,\beta_3$ computed from (\ref{18}) are
several times smaller than computed in $\overline{MS}$ scheme.

To compare $\alpha_R$ in Eqs. (\ref{16}), (\ref{17}) with the loop
expansion of SPT one can consider $Q^2$ large enough so that one
can use the asymptotics (\ref{15}) for $\psi(x)$   and keep the
leading (logarithmic) term. With the notation
\be
L\equiv \ln  \frac{Q^2+(M^{(0)}_R)^2}{\Lambda^2},~~
\alpha_R^{(0)}\equiv \frac{4\pi}{\beta_0L}, \label{19}\ee one
obtains the expansion in $1/L$ and $\frac{\ln L}{L}$, which looks
like \be\alpha_R=\alpha^{(0)}_R\left\{ 1-
\frac{2\beta_1}{\beta^2_0} \frac{\ln L}{L} +
\frac{4\beta^2_1}{\beta^4_0 L^2} \left[\left(\ln
L-\frac12\right)^2+b\right ]+O\left(\frac{\ln
L}{L}\right)^3\right\}.\label{20}\ee Here
$b=-\frac{\beta^2_0}{2\beta_1}
\left(\Delta-\frac12\right)-\frac14$, and for $\Delta=1.5$ one has
$b=-1.436$, while in the $\overline{MS}$ $b=0.26$ for $n_f=0$.
Thus the difference with SPT occurs only in the 3-loop result and
is not large.  Typically at $Q^2=3$ GeV$^2$ the exact expression
(\ref{16}) yields $\alpha_R=0.26$ while the 2-loop $\overline{MS}$
result is $\alpha_R=0.256 (n_f=0)$.

\section{Conclusions}

We conclude this paper with the discussion of the OPE for the case
of $e^+e^-$ annihilation which will stress the  new features of
the $\alpha_R(Q^2)$ behaviour (\ref{12a}). Expanding the
polarization function $\Pi(Q)$ in powers of $\alpha_s$ one has
\be
\Pi(Q) =\Pi^{(0)}(Q) +\alpha_s \Pi^{(1)} (Q)
+\alpha_s^2\Pi^{(2)}(Q)+...\label{21}\ee where $\Pi^{(0)}(Q)$ can
be computed as the spectral sum \cite{13} as follows
\be
\Pi^{(0)}(Q) =\frac{1}{12\pi^2}\sum
\frac{c_n}{(M^{(0)}_n)^2+Q^2}=-\frac{N_c}{12\pi^2}
\psi\left(\frac{Q^2+(M_n^{(0)})^2}{m^2}\right)\label{22}\ee and
similarly for the hybrid sums $\Pi^{(1)}, \Pi^{(2)}, ...$

This should be compared with the standard OPE  expansion \cite{14}
\be
\Pi(Q^2)=-\frac{N_c}{12\pi^2} \left( 1+\frac{\alpha_s}{\pi}\right)
\ln \frac{Q^2}{\mu^2} + \frac{6m^2_q}{Q^2}+\frac{2m_q \lan \bar q
q\ran}{Q^4} +\frac{\alpha_s\lan FF\ran}{12\pi
Q^4}+...\label{23}\ee
 where $m_q=\frac{m_n+m_d}{2}$. As was discussed in \cite{9,13},
 the leading asymptotic term in (\ref{22})  reproduces the
 partonic answer  $\frac{N_c}{12\pi^2}\ln \frac{Q^2}{\mu^2}$,
 supporting the idea of Quark-Hadron Duality (QHD).

 The next asymptotic terms of (\ref{22}) produce the  OPE terms
 $O\left(\frac{1}{Q^2}\right)$, $O\left(\frac{1}{Q^4}\right)$,
 etc. One usually imposes the QHD requirement, trying to reproduce
 the OPE coefficients in (\ref{23}), and in particular the absence
 of  $O\left(\frac{1}{Q^2}\right)$term for $m_q=0$ \cite{13}. Here
 we  notice, that the form (\ref{16}) produces additional power
 terms of  somewhat different structure.
Namely, taking the subleading asymptotic term in (\ref{15}) for
$\alpha_R$ one has
\be
\alpha_R=\alpha_R^{(0)} + \frac{m^2}{2(Q^2+m^2)}
\frac{\beta_0}{4\pi}  (\alpha^{(0)}_R)^2+...\label{24}\ee Assuming
that $\Pi^{(1)}(Q)$ has the same asymptotics  as $\Pi^{(0)}$ ( to
reproduce the term $\frac{\alpha_s}{\pi} \ln \frac{Q^2}{\mu^2}$ in
(\ref{23})) one obtains a new power term
\be
\Delta \Pi (Q^2) =- \frac{N_c}{12\pi^2} \frac{\alpha^{(0)}_R}{\pi}
\frac{m^2}{2Q^2}\label{25}\ee which is $O\left(\frac{1}{Q^2\ln
Q^2}\right)$ and has a negative sign.

Thus the meromorphic $\alpha_R$, Eq. (\ref{16}) produces the
power-like terms of new kind. Numerically these terms are smaller
than those originating  from asymptotic expansion of (\ref{22}),
and phenomenologically the appearance of $O(1/Q^2)$ terms is not
forbidden by experiment \cite{15}.

As a summary, a new type of solutions of RG equations is suggested
where $\alpha_R(Q)$ is a meromorphic function of $Q$ with poles
corresponding to the physical poles of two-gluon hybrid states.
The resulting $\alpha_R(Q)$ is finite for $Q^2>0$ and agrees well
with $\alpha_{\overline{MS}}(Q)$ for $Q>1$ GeV and is
phenomenologicaly acceptable for all positive $Q^2$.

The author is grateful to I.Ya.Arefieva, A.M.Badalian, A.L.Kataev,
N.V.Kras\-ni\-kov, A.A.Pivovarov, D.V.Shirkov, for useful remarks
and discussions. The work  is supported
  by the Federal Program of the Russian Ministry of Industry, Science, and Technology No 40.052.1.1.1112.\\

\end{document}